\documentclass[aps,prb,twocolumn,groupedaddress,nolongbibliography]{revtex4-2}

\usepackage{times}
\usepackage{amsmath}
\usepackage{amsfonts}
\usepackage{txfonts}
\usepackage[T1]{fontenc}
\usepackage[utf8]{inputenc}
\usepackage[english]{babel}
\usepackage{epsfig}
\usepackage{graphicx}
\usepackage{array,tabularx}	
\usepackage[colorlinks=true, linkcolor={blue}, citecolor={blue}, urlcolor={black}]{hyperref}

\begin{document}


\title{2D Li$^{\bf +}$ ionic hopping in Li$_{\bf 3}$InCl$_{\bf 6}$ as revealed by diffusion-induced nuclear spin relaxation}


\author{Florian Stainer}
\affiliation{Graz University of Technology, Institute of Chemistry and Technology of Materials (NAWI Graz), Stremayrgasse 9, 8010 Graz, Austria}

\author{H. Martin R. Wilkening}
\email[]{wilkening@tugraz.at}
\affiliation{Graz University of Technology, Institute of Chemistry and Technology of Materials
(NAWI Graz), Stremayrgasse 9, 8010 Graz, Austria}


\date{\today}

\begin{abstract}
Ternary Li halides, such as Li$_3$MeX$_6$ with, \textit{e.g.}, Me = In, Sc, Y and X = Cl, Br, are in the center of attention for battery applications as these materials might serve as ionic electrolytes. To fulfill their function, such electrolytes must have an extraordinarily high ionic Li$^+$ conductivity. Layer-structured Li$_3$InCl$_6$ represents such a candidate; understanding the origin of the rapid Li$^+$ exchange processes needs, however, further investigation. Spatially restricted, that is, low-dimensional particle diffusion might offer an explanation for fast ion dynamics. It is, however, challenging to provide evidence for 2D diffusion at the atomic scale when dealing with polycrystalline powder samples. Here, we used purely diffusion-induced $^7$Li nuclear magnetic spin relaxation to detect anomalies that unambiguously show that 2D Li diffusion is chiefly responsible for the dynamic processes in a Li$_3$InCl$_6$ powder sample. The change of the spin-lattice relaxation rate $1/T_1$ as a function of inverse temperature $1/T$ passes through a rate peak that is strictly following \textit{asymmetric} behavior. This feature is in excellent agreement with the model of P.\,M. Richards suggesting a logarithmic spectral density function $J$ to fully describe 2D diffusion. Hence, Li$_3$InCl$_6$ belongs to the very rare examples for which 2D Li$^+$ diffusion has been immaculately verified. We believe that such information help understand the dynamic features of ternary Li halides.
\end{abstract}


\maketitle

\section{\label{intro}Introduction}

Materials with low-dimensional features of different types \cite{geim,new,graph2,graph6,atlas,revv} play a prominent role in many areas of physics and materials science \cite{revv}. They might clear the way for new applications that ground on functional compounds whose crystal structures give rise to 1D or 2D electronic, magnetic, mechanical, optical, optoelectronic, chemical and even neuromorphic and biomedical properties \cite{revv}.

Restricting ourselves to electric transport properties \cite{harris}, in literature we find many studies focussing on the spatially confined movements of any kinds of electronic charge carriers in, \textit{e.g.}, inorganic and organic layer-structured compounds, that is, in 2D materials \cite{katz,ggr,2dg,nat,reve1,one,twot,three,four,allg}. However, the number of investigations accepting the challenge to unequivocally identify 2D transport of \textit{ionic} species is still much lower \cite{conr,kuch,eppl,stan4,eppiop,kuhnjacs,hiebl}. Many mixed conductors, \textit{i.e.}, compounds that show both electronic and ionic conductivity such as TiS$_2$ \cite{wilkprb}, graphite \cite{langerc} or Li$_x$CoO$_2$ ($0.5 < x \leq 1 $) \cite{sugic}, act as 2D host structures that can accept and release Li$^+$ in a highly reversible manner. The rapid insertion and facile de-insertion processes of small guest species without destroying the rigid host structure are at the heart of modern electrochemical energy storage systems \cite{Whitt}. Research on this insertion principle finally culminated in the advent of the now well known lithium-ion batteries for whose development M.\,S. Whittingham, J.\,B. Goodenough, and A.\,Yoshino were jointly awarded the Nobel prize in chemistry in 2019 \cite{nobel,mei}.

While some crystal structures provide classical van-der-Waals gaps for 2D diffusion, such as single phase TiS$_2$ \cite{wilkprb}, other materials offer geometric restrictions to guide the ions along the 1D or 2D transport pathways \cite{eppl,kuhnjacs,prutch3}. Examples include the hexagonal form of LiBH$_4$ showing 2D Li$^+$ diffusion \cite{eppl} or the binary compound Li$_{12}$Si$_{7}$ in which the Li$^+$ ions quickly hop along the stacked, five-membered Si$_5$ rings, thus performing quasi 1D diffusion \cite{kuhnjacs}.

The low-dimensional diffusion pathways of the two examples mentioned have been studied by nuclear spin relaxation (NSR) \cite{eppl,kuhnjacs}, which is highly suited to probe the internal magnetic dipolar or electric quadrupolar field fluctuations caused by the diffusive hopping processes of the spin-carrying atoms or ions \cite{Wilkening2012,hogr}. Most importantly, while classical orientation-dependent conductivity measurements require single crystals to probe the anisotropy of the transport process, for NSR measurements powder samples are, however, sufficient. NSR probes a spectral density function $J$ that contains the information about the temporal interactions to which the spins are subjected \cite{wilkprb,Heitjans2005}; the underlying basics of NSR measurements are provided in the Supporting Information. Besides information on activation energies and Arrhenius pre-factors determining a classical over-barrier hopping process, the shape of the NSR rate peak does also depend on the dimensionality of the diffusive motions of the spins \cite{eppl,Wilkening2012}. The powerfulness of this approach to identify low-dimensional particle movements has recently been demonstrated on polycrystalline $\beta$-Li$_3$PS$_4$ \cite{prutch3}. With the help of NSR measurements we were able to reveal even buried low-dimensional pathways in this thiophosphate that gives, at first glance, no occasion for 1D (zig-zag) diffusion pathways \cite{prutch3}.

\begin{figure*}[t!]
\includegraphics{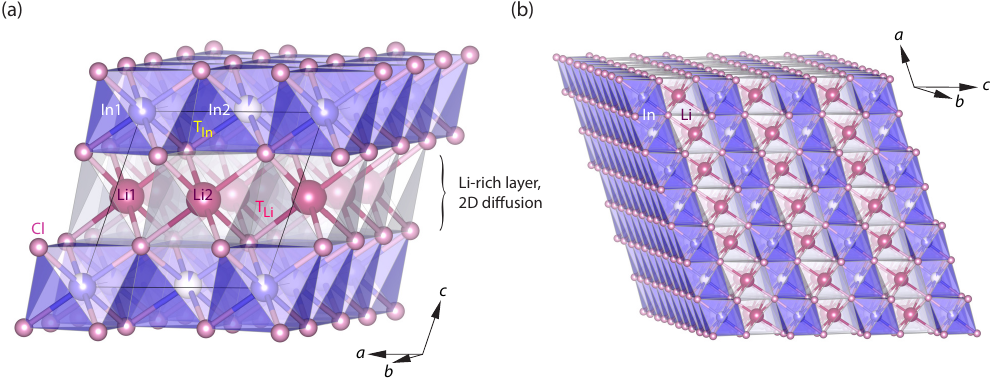}
\caption{(a) Crystal structure of monoclinic Li$_3$InCl$_6$ adopting $C{\textrm{2}}/m$ symmetry; here, it is drawn such that Li1 on 4$h$ and Li2 residing on $2d$ fully occupy the interlayer void between the In-rich layers, as proposed by single X-ray diffraction analysis by G. Meyer and co-workers. (b) Illustration to better visualize the (ordered) layered structure of this ternary halide. Recently, Helm et al. suggested that Li does also regularly occupy the tetrahedral void indicated as T$_{\textrm{Li}}$. Moreover, they suggested that a relatively large number fraction of Li ions does also reside on the In2 position, which is here either not occupied by Li$^+$ or to a much lesser extent.}
\label{fig1}
\end{figure*}

The compounds analyzed in this way so far, exhibit rather rapid 2D Li$^+$ diffusion. Hence, enabling 2D ionic transport can be regarded as a valid design principle to find materials that can serve as powerful solid electrolytes in devices for electrochemical energy storage. Recently, so-called ternary halides \cite{s34,kwa,tuo,ma2,xia3} have re-aroused the interest of workgroups searching for fast ionic conductors. Lithium indium chloride, Li$_3$InCl$_6$ (see Figure~\ref{fig1}), represents one of the most fascinating representatives of this class of materials as it provides geometric features which could indeed lead to facile 2D Li$^+$ diffusion between the In-rich layers \cite{Lutz,meyer,LiLi_2}. Explaining its conductivity properties with the help of dimensionality effects would definitely contribute to disclosing the structural principles that lead to fast ion transport in these solids.

Here, we prepared a phase-pure, polycrystalline Li$_3$InCl$_6$ sample via wet-chemical synthesis and studied the diffusion-induced $^7$Li NSR rates $1/T_1$ as a function of temperature. Provided data from highly accurate measurements are at hand, the rates can only be understood in terms of 2D diffusion fully confirming the semi-empirical 2D spectral density function P.\,M. Richards has introduced in 1978 \cite{rich,Gombotz2019}.

\section{\label{exp}Experiment}

Li$_3$InCl$_6$ has been prepared through a water-assisted dissolution-precipitation route \cite{LiLi}. For that purpose, the mixture of the starting materials (2~g) was dissolved in 10~mL of miliQ water. The clear solution was transferred to a Schlenk flask and the water was evaporated. The remaining white solid was dried at 100~°C under vacuum for a period of 2~hours resulting in the hydrated precursor Li$_3$InCl$_6$\,$\cdot$\,$x$H$_2$O. In a second 4-h drying step carried out at $\vartheta =$ 200~°C, we obtained the final polycrystalline product Li$_3$InCl$_6$. After having removed any residual water through this final temperature treatment, the product was handled strictly under Ar Atmosphere or vacuum to avoid any further contamination with water.

The final product has been characterized by X-ray powder diffraction. X-ray powder patterns were recorded on a Rigaku MiniFlex 600 powder diffractometer operating with Bragg Brentano geometry and Cu-K$\alpha$-radiation. We covered a 2$\theta$ range of 10° $\leq$ 2$\theta$ $\leq$ 90° (2.5 °\,min$^{-1}$, step size 0.1°). During the measurements, the sample has always been kept under inert gas atmosphere (Argon) using an airtight sample holder constructed by Rigaku.

$^7$Li NMR spin-lattice relaxation rates were recorded at $\omega_0/2\pi = 116~\rm MHz$ and 194~MHz using Bruker Avance NMR spectrometers equipped with ceramic broadband probe heads (Bruker) that allowed us to adjust the temperature (Eurotherm controller) inside the sample chamber with a stream of heated nitrogen gas \cite{Epp2013}. We used the saturation recovery pulse sequence to monitor the buildup curves of the longitudinal magnetization as a function of waiting time. The 90° excitation pulse that perturbs the spin population accruing to the Boltzmann term ranged from 2 to 2.5~$\mu$s depending on temperature $T$. These transients follow single exponential time behavior and were analyzed to extract the diffusion-induced rate $1/T_1$, which is proportional to the spectral density function $J(\omega_0)$ that has components at $\omega_0$ \cite{wilkprb,Wilkening2012}, see Supporting Information. Here, diffusion-induced relaxation towards thermodynamic spin equilibrium is assumed to be controlled by both magnetic dipolar and electric quadrupolar interactions as $^7$Li is a spin-3/2 nucleus. Independent of the exact nature of spin-lattice relaxation, the shape of $J$ allows for the determination of the dimensionality of the diffusion process, see Supporting Information containing an introduction into the basics behind the method.

\section{\label{rd}Results and Discussion}

\begin{figure*}
\includegraphics{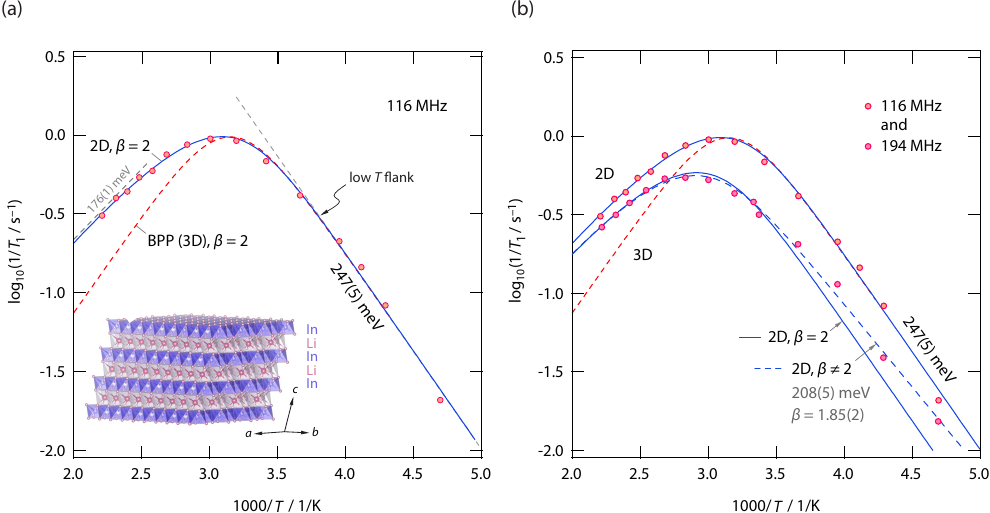}
 \caption{\label{fig2} (a) Diffusion-induced $^7$Li NMR spin-lattice relaxation rates of Li$_3$InCl$_6$ analyzed in the frame of an Arrhenius plot. The rates were recorded at 116~MHz. At this frequency the low-temperature slope points to an activation energy of 247(5)~meV. While the solid line shows a fit according to Richards NSR model for 2D diffusion, the dashed line indicates the response that would be expected for 3D diffusion. $\beta =2$ reflects pure BPP-type relaxation behavior in the limit $\omega_0\tau_c \gg 1$ indicating random instead of correlated ion diffusion. The asymmetric shape of the Richards peak obtained is diagnostic for low-dimensional diffusion. In the 2D relaxation model the formal activation energy of 176~meV has no physical parallel as it does not correspond to a real barrier. This reduced slope (71\% $\approx 3/4\,E_{\rm a}$) is a consequence of the spectral density function $J$ of 2D diffusion; for 1D diffusion it would be reduced to $E_{\rm a}/2$, see Supporting Information (Table S1). (b) NMR relaxation rate peaks of Li$_3$InCl$_6$ recorded at two different Larmor frequencies as indicated. Here, 2D diffusion produces a dispersive, frequency-dependent regime governing the rates of the high-$T$ slope. See text for further details}
\end{figure*}

In Figure~\ref{fig2}\,(a) we plotted the purely diffusion-induced $^7$Li NSR rates $1/T_1$ of Li$_3$InCl$_6$ as a function of the inverse temperature using an Arrhenius representation. Starting from low temperatures, the rates follow a linear increase that is characterized by an activation energy $E_{\rm a}$ of 247(5)~meV. In this so-called low-temperature regime, where we have to deal with $\omega_0\tau_{\rm c} \gg 1$, the NMR rate is, in general, sensitive to local jump events of the Li$^+$ ions in Li$_3$InCl$_6$. Hence, we identify this activation energy as that describing the \textit{elementary} steps of Li$^+$ diffusion in the ternary halide.

Increasing the temperature even more, the rates pass through a maximum located at $T$ $\approx$ 320~K. At this temperature, the mean residence time $\tau$ of the Li ions equals the correlation time $\tau_c$, which is given by $\omega_0\tau_{\rm c} \approx 1$, that is, determined by the angular Larmor frequency, $\omega_0 \approx 7.28 \times 10^8$~rad\,s$^{-1}$ used to sample the rates. The jump rate $\tau^{-1}$ is assumed to depend on temperature following the Arrhenius law with $k_{\rm B}$ denoting Boltzmann's constant:

\begin{equation}
\tau^{-1} \approx \tau_{\rm c}^{-1} = \tau_{\rm c0}^{-1} \exp(-E_{\rm a}/(k_{\rm B}T)).
\end{equation}

Importantly, for 3D diffusion we would expect the $1/T_1$ NSR rates of the peak shown in Figure~\ref{fig2}\,(a) to pass into a high-$T$ flank that is governed by a slope either equal or larger than that probed on the low-$T$ side. Here, however, the diffusion-induced $^7$Li NSR rates $1/T_1$ reveal a weaker-than-expected temperature dependence in this $T$ regime. Such a behavior shown in Figure~\ref{fig2}\,(a) can only be explained by assuming a low-dimensional diffusion process. Such a process is also suggested by the crystal structure of Li$_3$InCl$_6$ (Figure~\ref{fig1}), which favors a 2D process rather than a 3D one that would be responsible for the (symmetric) $1/T_1(1/T)$ NMR peak.

To corroborate this assumption, we used the spectral density function $J(\omega_{0})$ introduced by Richards \cite{rich} for 2D diffusion to analyze our data; a brief introduction into the basics of NMR relaxation measurements is given in the Supporting Information. The solid line in Figure~\ref{fig2}\,(a) shows a fit according to the following relationship which takes into account a logarithmic frequency dependence of the rate in the high-$T$ limit ($\omega_0\tau_{\rm c} \gg 1$); the constant $C''_{0}$ is proportional to the effective coupling constant and determines the height of the whole rate peak $1/T_1(1/T)$.

\begin{equation}
J(\omega_{0}) \propto 1/T_{1} = C''_{0} \dfrac{\tau_{\rm c}}{1+(\omega_{0}\tau_{\rm c})^\beta}
\end{equation}

This dependence causes the NMR rate peak to assume an asymmetric shape as also shown in Figure~\ref{fig2}\,(a). The excellent agreement between experimental data and the semi-empirical Richards model \cite{rich} proves that Li$^+$ diffusion in Li$_3$InCl$_6$ is of 2D nature. For comparison, the conventional response for 3D diffusion \cite{eppl} is also shown, see the dashed line in Figure~\ref{fig2}\,(a). The clear deviation of the two models in the high-$T$ limit made it possible to unequivocally identify Li$_3$InCl$_6$ as a 2D ionic conductor.

In both cases, and especially for the curve representing 2D diffusion, the best fit is obtained for $\beta =2$. This parameter, if adopting values smaller than 2, reduces the slope on the low-$T$ side as illustrated in the Supporting Information. $\beta < 2$ \cite{kuch} is usually used to take into account correlation effects \cite{Heitjans2005,hogr} leading to a sub-quadratic frequency dependence in the limit $\omega_0\tau_{\rm c} \gg 1$ which also manifests in a clear change of the slope in the limit $\omega_0\tau_{\rm c} \gg 1$ (see Figure~S1\,(a), Supporting Information).

In the present case, a classical quadratic frequency dependence, $1/T_1 \propto \omega_0^{-2}$, which follows the idea of the pioneering model of Bloembergen, Purcell and Pound \cite{BPP1948} is fully sufficient to precisely mirror the NMR response at least at 116~MHz (Figure~\ref{fig2}\,(a)). $\beta =2$ does not point to strong correlation effects between the jumping ions which would lead to a lower slope and thus to a lower (apparent) activation energy. Here, we interpret the value of 247(5)~meV as the elementary barrier the ions need to surmount to diffuse within the Li-rich layer. In contrast, the same slope should be seen in the high-$T$ region for the hypothetic case of 3D diffusion, as indicated by the dashed line in Figure~\ref{fig2}\,(a). Such a BPP-type relaxation behavior seen within Richards 2D model \cite{rich} suggests that we have to deal with a symmetric potential landscape in which the ions perform random walks rather than correlated motions. In such a uniform landscape there will be no difference between short-range and long-range ion diffusion as earlier discussed also for another 2D ion donators \textit{viz.} Li$_x$TiS$_2$.

As mentioned above, Richards's 2D spectral density function predicts a logarithmic frequency dependence of $1/T_1$ on the high-$T$ side (see also Supporting Information, Figure~S1\,(b)). Such a behavior is in contrast to what is expected for 3D diffusion, for which we have $1/T_1 \neq \omega_0$ approaching the limit $\omega_0\tau_{\rm c} \ll 1$. Indeed, repeating the $^7$Li NMR spin-lattice relaxation measurements at a higher Larmor frequency $(\omega_0/2\pi = 194~\rm MHz)$ reveals a frequency dependence of $1/T_1$ on the high-temperature side of the diffusion-induced peak (\textit{cf.}~see Figure~\ref{fig2}\,(b)). This feature perfectly corroborates the 2D dynamic situation in Li$_3$InCl$_6$. In agreement with the maximum condition, $\omega_0\tau_{\rm c} \approx 1$, the rate peak recorded at higher magnetic field is slightly shifted towards higher temperature, see Figure~\ref{fig2}\,(b).

We have to note that the NMR measurements carried out at 194~MHz (see Figure~\ref{fig2}\,(b)) reveal a lower slope on the low-$T$ side than expected. The activation energy reduces from 247~meV to only 208~meV; $\beta$ changes from 2 to 1.85. Obviously, at increased Larmor frequency, that is, at a shorter time window, the effect of correlation effects such as (localized) forward-backward jump processes begin to influence the slope of the NMR rate peak in this temperature limit.

Finally, let us briefly note that previous studies \cite{maas} also attempted to tackle the same problem, that is, interpreting NSR rates in terms of 3D or 2D jump diffusion. However, certain difficulties did not result in such a clear picture as is obtained here. In the present case, we believe that the high purity of the crystalline powder sample, for which we also assume a poor defect density, lead to NMR rates solely induced by the 2D nature of the self-diffusion process. As a side note, purely diffusion-induced $1/T_1$ rates are an imperative necessity to precisely uncover the 2D dynamic features of the ternary halide.

\section{\label{con}Conclusion}

Li$_3$InCl$_6$, crystallizing with a layered structure, turned out to be an optimal model system to examine the Richards model for its suitability to describe spatially constrained 2D ionic dynamics in a crystalline matrix. The purely diffusion-induced $^7$Li NMR relaxation rates point to an activation energy for Li$^+$ displacements of 247 meV and fully agree with the Richards's model introduced to describe 2D hopping diffusion. This model predicts an asymmetric $1/T_1(1/T)$ rate peak and a logarithmic frequency dependence of the rates in the high-$T$ limit. Our NSR results build the experimental reflection of this semi-empirical model. Low dimensional Li$^+$ diffusion is believed to be key for the understanding of the superior transport properties of materials related with Li$_3$InCl$_6$.

\begin{acknowledgments}
Financial support by the DFG in the frame of the former research unit molife (FOR1277) is greatly acknowledged.
\end{acknowledgments}

\bibliography{lit_stainer}

\end{document}